\documentclass{emulateapj}
\usepackage{apjfonts}
\lefthead{HAN} 
\righthead{MICROLENSING BINARY PROPERTIES}

\newcommand{\te}{t_{\rm E}}

\newcommand{\thetae}{\theta_{\rm E}}

\def\eqalign#1{\null\,\vcenter{\openup\jot
        \ialign{\strut\hfil$\displaystyle{##}$&$
        \displaystyle{{}##}$\hfil \crcr#1\crcr}}\,}

\begin{document}
\title{Investigating Binary Properties with Next-Generation 
Microlensing Surveys}

\author{Cheongho Han}
\affil{
Program of Brain Korea 21,
Institute for Basic Science Research, 
Department of Physics,\\
Chungbuk National University, Chongju 361-763, Korea;
cheongho@astroph.chungbuk.ac.kr}



\begin{abstract}
We explore the usefulness of future gravitational microlensing surveys 
in the study of binary properties such as the binary fraction and the 
distributions of binary separation and mass ratio by using the binary 
sample detectable through a channel of repeating events.  For this, we 
estimate the rate of repeating microlensing eventstoward the Galactic 
bulge field based on standard 
models of dynamical and physical distributions of Galactic matter combined 
with models of binary separation and mass function.  From this, we find 
that the total number of repeating events expected to be detected from 
$\sim 4$-year space-based surveys will be $\sim 200$--400, that is 
$\sim 40$--50 times higher than the rate of current surveys.  We find that 
the high detection rate is due to the greatly improved sensitivity to 
events associated with faint source stars and low-magnification events.  
We find that the separation range of the binaries to be covered by the 
repeating events will extend up to 100 AU.  Therefore, the future lensing 
surveys will provide a homogeneous sample that will allow to investigate 
the statistical properties of Galactic binaries unbiased by brightness 
of the binary components.
\end{abstract}

\keywords{binaries: general -- gravitational lensing}

\section{Introduction}

Massive searches for gravitational microlensing events have been and 
are being carried out by monitoring a large number of stars located 
in the Galactic bulge and Magellanic Clouds \citep{alcock93, aubourg93, 
udalski93}.  Current lensing surveys are detecting events with a rate 
of $\sim 500$ events per year, most of them toward the Galactic bulge 
field, and the total number of detected events now exceeds 2000.  Among 
them, a significant fraction of the events are caused by binaries.

Although a sizable number of binary lensing events have been detected,
there have been few investigations about the properties of binaries 
like the binary fraction and the distributions of the binary separation 
and mass ratio based on the binary event sample.  There are several 
reasons for this.  First, current lensing detection is strongly biased 
for binaries with separations equivalent to the Einstein radius of the 
combined mass of the binary, and thus the range of binary separation 
covered by lensing is very narrow.  Second, for events detected through 
this {\it resonant} channel, it is difficult to quantify the detection 
efficiency due to the non-linear lensing behavior.  Third, interpreting 
the lightcurve of these events suffers from degeneracy of parameters 
and thus accurate determinations of the separation and mass ratio are 
difficult \citep{dominik99}.  As a result, there has been no meaningful 
result about the statistical properties of binaries although there have 
been several reports of binary event samples \citep{alcock00, jaroszynski02, 
jaroszynski04}.

Besides the resonant channel, binaries can be detected through another 
channel of repeating events \citep{distefano96}.  Binary detection 
through this channel is possible if an event is produced by a widely 
separated binary lens and the source trajectory approaches both lens 
components.  The resulting lightcurve is approximated by the superposition 
of the single lensing events caused by the individual binary components.  
Using the sample of binaries detected through this channel has various 
advantages in the studies of binary properties.  First, being able to 
detect wide separation binaries, the separation range of binaries that 
can be studied with microlensing can be greatly expanded.  Second, the 
detection efficiency is estimated in a straightforward manner and thus 
can be easily quantified.  Third, determinations of the binary lens 
parameters do not suffer from degeneracy and thus the parameters can 
be accurately determined.

Despite these advantages, current lensing surveys are very inefficient
in detecting binary events through the repeating channel.  This is 
because repeating events require special lens-source geometry and 
thus they are very rare.  \citet{distefano96} pointed out that the 
detection rate of repeating binary lensing events could be increased 
by conducting follow-up observations of events.  However, these 
follow-ups require long-term observations of the individual events.  
Considering that there are limited resources of follow-up observations 
and the main purpose of the follow-up observations currently being 
conducted is searching for extrasolar planets \citep{bond02, park04, 
cassan04}, implementing the proposed follow-up observations for the 
detections of wide binaries is practically very difficult.

Recently, serious discussions are going on about future lensing surveys 
with the prime goal of searching for a large sample of extrasolar planets 
including Earth-mass planets.  The {\it Microlensing Planet Finder} 
({\it MPF}), which succeeded the original concept of the {\it Galactic 
Exoplanet Survey Telescope} ({\it GEST}) \citep{bennett02}, is a space 
mission exclusive for microlensing and it will be equipped with a 1.1 m 
aperture telescope, which images a 1.3 ${\rm deg}^2$ field of view.  
The `Earth-Hunter' project is a ground-based survey that plans to achieve 
$\sim 10$ minute sampling by using a network of three 2 m class wide 
field ($\sim 4\ {\rm deg}^2$) telescopes scattered over the southern 
hemisphere (A.\ Gould, private communication).  These next-generation 
surveys will have the capability of continuously monitoring stars at 
high cadence by using very large format imaging cameras and thus dispense 
with the need of follow-up observations.

In this paper, we point out the usefulness of the future lensing 
surveys in the study of binary properties.  We demonstrate that 
a large number of binary events are expected to be detected in the 
future lensing surveys through the channel of repeating events thanks 
to the greatly improved sensitivity to events associated with faint 
source stars and low-magnification events.

The paper is organized as follows.  In \S\ 2, we briefly describe the 
basics of repeating binary lensing events.  In \S\ 3, we estimate the 
detection rate of repeating binary events expected from future lensing 
surveys based on standard models of the physical and dynamical 
distributions of Galactic matter combined with models of binary 
separation and mass function under realistic observational conditions.  
We also investigate the characteristics of the events.  In \S\ 4, we 
discuss about the advantages of using the repeating binary event sample
in the studies of binary properties.  In \S\ 5, we summarize 
the results and conclude.

\section{Repeating Events}

If a lensing event is caused by a binary with a projected separation 
between the lens components significantly larger than the Einstein 
radius of the combined mass of the binary, the individual binary 
components behave as if they are independent single lenses.  Then, 
the resulting lightcurve is well represented by the superposition of 
the single lensing events, i.e., 
\begin{equation}
A \simeq A_1 + A_2 - 1,
\label{eq1}
\end{equation}
where $A_1$ and $A_2$ are the magnifications of the single lensing events 
caused by the individual binary components.  We use the subscripts `1' 
and `2' to designate the events caused by the individual binary 
components according to the time of the source's approach and refer 
the individual single lensing events as the {\it first} and {\it second} 
events, respectively.  The Einstein radius is related to the mass of 
the lens, $M$, and distance to the lens, $D_{\rm L}$, and source, 
$D_{\rm S}$, by
\begin{equation}
\eqalign{
\thetae = &  \sqrt{4GM\over c^2} \left({1\over D_{\rm L}}-
             {1\over D_{\rm S}} \right)^{1/2} \cr
\sim      &  550\ 
             {\rm \mu as}\ \left( {M\over 0.3\ M_\odot}\right)^{1/2}
             \left( {D_{\rm S}\over 8\ {\rm kpc}}\right)^{-1/2}
             \left( {D_{\rm S}\over D_{\rm L}}-1\right)^{1/2}.  \cr
}
\label{eq2}
\end{equation}
For each single lens event, the magnification is related to the lensing 
parameters by
\begin{equation}
A = {u^2+2 \over u (u^2+4)^{1/2}},\qquad
u = \left[\left( {t-t_0 \over t_{\rm E}} \right) 
+ u_0^2\right]^{1/2},
\label{eq3}
\end{equation}
where $u$ is the lens-source separation normalized by the Einstein radius 
of each lens component, $t_{\rm E}$ is the timescale required for the 
source to transit the Einstein ring radius (Einstein timescale), $t_0$ is 
the time of the source star's closest approaches to the lens, and $u_0$ 
is the lens-source separations at that moment (impact parameters) normalized 
by the Einstein radius.  The Einstein radius of each binary component is 
related to that of the combined mass by $\theta_{{\rm E},i} =\sqrt{m_i} 
\thetae$, where $m_i=M_i/(M_1+M_2)$ is the mass fraction of the binary 
component.


The condition for a wide separation binary to cause a repeating event 
is that the source trajectory closely approaches both lens components.  
If $d$ is the angular separation between the lens components and $b_{1}$ 
and $b_{2}$ represent the impact parameters (also in angular units) of 
the source trajectory to the individual lens components, then this 
condition is represented by
\begin{equation}
{b_{{\rm th},1} - b_{{\rm th},2} \over d} 
\leq
 \sin\alpha
\leq
{b_{{\rm th},1} + b_{{\rm th},2} \over d},
\label{eq4}
\end{equation}
where $\alpha$ denotes the orientation angle of the source trajectory 
with respect to the binary axis and $b_{{\rm th},1}$ and $b_{{\rm th},2}$ 
are the maximum values of the impact parameters for the detection of the 
single-lensing events involved with the individual binary components.
By normalizing the binary separation in terms of the Einstein radius 
of the combined mass of the binary, $s=d/\theta_{\rm E}$, and the 
impact parameters in terms of the Einstein radii of the individual 
lens components, $u_{0,{\rm th},i}=b_{{\rm th},i}/\theta_{{\rm E},i}$,
equation~(\ref{eq4}) is expressed as 
\begin{equation}
{\sqrt{m_1}u_{0,{\rm th},1} - \sqrt{m_2}u_{0,{\rm th},2} \over s} 
\leq
 \sin\alpha
\leq
{\sqrt{m_1}u_{0,{\rm th},1} + \sqrt{m_2}u_{0,{\rm th},2} \over s}.
\label{eq5}
\end{equation}

\section{Detection Rates of Repeating Events}

In this section, we estimate the rate of repeating binary lensing 
events expected to be detected by future lensing surveys and compare 
the rate to that of current surveys.  For this, we conduct simulations 
of repeating events.

\subsection{Lensing Event Simulation}

With the 
standard models of 
dynamical and physical distributions of Galactic matter combined with 
models of binary separation and mass function, we produce repeating 
events toward the Galactic bulge field as follows.

First, survey is assumed to be conducted toward the Galactic bulge 
field and the locations of the source and lens are allocated based on 
the mass distribution model of \citet{han03}.  In the model, the 
physical distribution of bulge matter is based on the deprojected 
infrared light density profile of \citet{dwek95}, specifically model 
G2 with $R_{\rm max}=5$ kpc from their Table 2.  The disk matter 
distribution is modeled by a double-exponential disk, which is 
expressed as 
\begin{equation}
\rho(R,z) =\rho_0 
\exp \left[-\left( {r-R_0\over h_R}+{|z|\over h_z}\right)\right], 
\label{eq6}
\end{equation}
where $(R,z)$ is the Galactocentric cylindrical coordinates, $R_0=8$ 
kpc is the distance of the Sun from the Galactic center, $\rho_0=
0.06\ M_\odot \ {\rm pc}^{-3}$ is the mass density in the solar 
neighborhood, and $h_R=3.5$ kpc and $h_z=325$ pc are the radial and 
vertical scale heights, respectively.

Second, the velocity distribution of the bulge is deduced from the 
tensor virial theorem, while the disk velocity distribution is assumed
to have a flat rotation speed of $v_c=220\ {\rm km}\ {\rm s}^{-1}$ and 
velocity dispersions along and normal to the disk plane of 
$\sigma_\parallel=30\ {\rm km} \ {\rm s}^{-1}$ and $\sigma_\perp=
20\ {\rm km}\ {\rm s}^{-1}$, respectively.  The resulting distribution 
of the lens-source transverse velocity from the combinations of the 
bulge and disk velocities is listed in Table 1 of \citet{han95}, 
specifically non-rotating barred bulge model.

Third, the absolute brightnesses of the Galactic bulge source stars are 
assigned on the basis of the luminosity function of \citet{holtzman98}, 
that was constructed from {\it Hubble Space Telescope} observations.  
Once the absolute magnitude is assigned, the apparent magnitude is 
determined considering the distance to the source star and extinction.  
The extinction is determined such that the source star flux decreases 
exponentially with the increase of the dust column density.  The dust 
column density is computed on the basis of an exponential dust 
distribution model with a scale height of $h_z=120\ {\rm pc}$, i.e.\ 
$\propto \exp(-|z|/h_z)$.  Based on the extinction determination of 
$A_V=1.28$ by \citet{holtzman98} and the mean ratio of extinction to 
color excess of $\langle R_{VI}\rangle=\langle A_V/E(V-I)\rangle=0.964$ 
determined by \citet{sumi04}, we normalize the amount of $I$-band 
extinction so that $A_I=0.49$ toward the Baade's Window.


Fourth, the mass of the lens is assigned based on the model mass 
function of \cite{gould00}, which is composed of stars, brown dwarfs 
(BDs), and stellar remnants of white dwarfs (WDs), neutron stars (NSs), 
and black holes (BHs).  The model is constructed under the assumption 
that bulge stars formed initially according to a double power-law 
distribution of the form
\begin{equation}
{dN\over dM} = k \left( {M\over 0.7\ M_\odot}\right)^\gamma,
\label{eq7}
\end{equation}
where $\gamma=-2.0$ for $M\geq 0.7\ M_\odot$ and $\gamma=-1.3$ for 
$M<0.7\ M_\odot$.  Based on this initial mass function, remnants 
are modeled by assuming that the stars with initial masses 
$1\ M_\odot< M <8\ M_\odot$, $8\ M_\odot < M < 40\ M_\odot$, and 
$m>40\ M_\odot$ have evolved into WDs (with a mean mass $\langle 
M\rangle\sim 0.6\ M_\odot$), NSs (with $\langle M\rangle\sim 1.35
\ M_\odot$), and BHs (with $\langle M\rangle\sim 5\ M_\odot$), 
respectively.  Then, the resulting mass fractions of the individual 
lens components are ${\rm stars}:{\rm BD}:{\rm WD}:{\rm NS}:{\rm BH}
=62:7:22:6:3$.

Once the first event is produced, we then introduce a second event.  
Following \citet{abt83}, we model the binary separation is uniformly 
distributed in $\log s$, i.e.\ $f(\log s)\equiv {\rm const}$.  To 
ensure that the part of the lightcurve induced by each of the individual 
lenses can be well approximated by a single lens lightcurve, the binary 
separation is restricted to be larger than three times of the combined 
Einstein ring radius, i.e.\ $s>3$.  For the mass function of the binary, 
we test three models.  In the first model, we assume that the two masses 
of the binary components are drawn independently from the same mass 
function as that of single stars.  This model is natural results of 
binary formation where binaries are formed through interactions between 
protostellar disks \citep{pringle89} or some other forms of capture.  
We refer this model as the `capture' model.  Other possible mechanisms 
of binary formation are fission of a single star and fragmentation of 
a collapsing objects.  Numerical calculations suggest that the former 
process results in a mass ratio distribution peaking at around $q=0.2$ 
\citep{lucy77}, while the latter results in more equal masses 
\citep{norman78}.  We refer these models as the `fission' and 
`fragmentation' models, respectively.  In the fission model, we model 
the mass ratio distribution as
\begin{equation}
f(q)=
\cases{
5q           &  for $q\leq 0.2$ \cr
-1.25q+1.25  &  for $q>0.2$,    \cr
}
\label{eq8}
\end{equation}
while the distribution of the fragmentation model is 
modeled as
\begin{equation}
f(q) = q.
\label{eq9}
\end{equation}
In the fission model, we set the total mass of the binary to be the same 
as that of a single star because a single star is split into two components 
of a binary in this model.  Then, the total mass of the binary is drawn 
from the mass function of single stars and the masses of the individual 
binary components are deduced from the mass ratio.  In the fragmentation 
model, on the other hand, the mass of the primary is drawn from the single 
star mass function and the mass of the companion is determined from the 
mass ratio.  As a result, the average mass of the binaries in this model 
is heavier than that of single stars.

\begin{deluxetable}{lll}
\tablecaption{Observational Conditions \label{table:one}}
\tablewidth{0pt}
\tablehead{
\colhead{specification} &
\colhead{current survey} &
\colhead{future survey} }
\startdata
platform                        & ground based               & space mission             \\
center of field                 & $(2.0^\circ,-2.7^\circ)$   & $(1.2^\circ,-2.4^\circ)$  \\
field of view                   & $21\ {\rm deg}^2$          & $2.6\ {\rm deg}^2$        \\
photometric precision           & $5\%$ at $I=19$            & $1\%$ at $I=21.5$         \\
sampling frequency              & 1 time/night               & 96 times/night            \\
fraction of clear nights        & $43\%$                & 100\%                     \\
observation duration per year   & 7 months                   & 9 months                  \\
\enddata
\tablecomments{ 
Summary of the observational conditions of the current and next-generation 
microlensing surveys.
}
\end{deluxetable}

\subsection{Observational Condition and Detection Criteria}

Event rate is not an absolute quantity but is subject to the 
conditions of observation and detection criteria.  We estimate the 
rate of repeating events by assuming the following observational 
conditions of the current and future lensing surveys.  We note that 
these conditions are based on those of the OGLE and {\it MPF} surveys, 
respectively.

The details of the observational conditions are as follows.
\begin{enumerate}
\item
For the observation fields, we assume that the current lensing 
survey is monitoring toward a bulge field of $(l,b) \sim (2.0^\circ,
-2.7^\circ)$ with a field of view of $(\Delta l,\Delta b)\sim
(3.0^\circ,7.0^\circ)$.  The future survey will monitor toward a 
denser field of $(l,b)\sim (1.2^\circ, -2.4^\circ)$ but with a
narrower field of view of $(\Delta l, \Delta b)\sim(0.93^\circ,
2.8^\circ)$.
\item
For the sampling frequency, we assume that the current survey is 
monitoring the same field with a sampling interval of 1 day, but the 
observation is interrupted by sporadic bad weather.  The nights of 
bad weather comprise $\sim 43\%$ of a total 7-month period of the 
bulge season.  The future lensing survey will continuously observe 
the target field with a sampling interval of 15 minutes except a 
3-month period to avoid the Sun.
\item
Despite that a star is observed by telescopes with similar apertures, 
the photometric precisions from ground and space-based observations 
are substantially different.  This is because the bulge field is very 
crowded and thus ground-based observations suffer from severe blending 
that contributes to the photometric noise.  To account for blending, 
we set the photometric precisions as $5\%$ at $I=19$ for the current 
survey and $1\%$ at $I=21.5$ for the future survey.
\end{enumerate}
We summarize the observational conditions in Table~\ref{table:one}.

\begin{figure}[t]
\epsscale{1.2}
\plotone{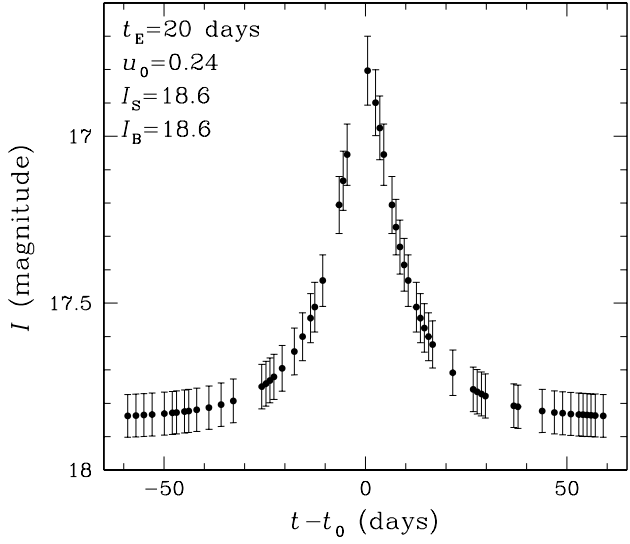}
\caption{\label{fig:one}
Lightcurve of an example lensing event that barely passes the imposed 
detection criteria.  See \S\ 3.2, for details of the criteria.
}\end{figure}

We judge the detectability of an event based on the uncertainties of 
the lensing parameters recovered from the lightcurves.  The uncertainties 
are estimated by computing the curvature matrix of $\chi^2$ surface.  
For the case of a lensing lightcurve, the curvature matrix is defined as 
\begin{equation}
b_{ij}=\sum_{k}^{N_{\rm obs}}
{\partial F_k\over \partial p_i}{\partial F_k\over \partial p_j}
{1\over \sigma_k^2},
\label{eq10}
\end{equation}
where $N_{\rm obs}$ is the number of observations, $F(t)=A(t)F_{\rm S}
+F_{\rm b}$ is the observed flux, $F_{\rm S}$ and $F_{\rm b}$ are 
fluxes of the unmagnified source and blended light, $\sigma_k$ is the 
photometric precision of each measurement, and $p_i\equiv (F_{\rm S}, 
F_{\rm b}, u_0, t_0, \te)$ are the five lensing parameters required 
to fit the lightcurve of a standard point-source single-lens event.  
Then, the uncertainties of the individual lensing parameters correspond 
to the diagonal components of the inverse curvature matrix (covariance 
matrix), i.e.
\begin{equation}
\sigma_{p_i} = \sqrt{c_{ii}};\qquad c\equiv b^{-1}.
\label{eq11}
\end{equation}
Among the lensing 
parameters, the Einstein timescale is of interest because it is the 
only lensing parameter related to the physical quantities of the lens 
system.  The timescale is also important because the mass ratio between 
the binary lens components of a repeating event is determined by $q=
(t_{{\rm E}, 2}/t_{{\rm E},1})^2$.  With the uncertainty of the timescale 
$\sigma_{t_{\rm E}}$, we set the criteria of event detection as 
\begin{equation}
\cases{
{\rm detection}     & if $\sigma_{t_{\rm E}}/\te \leq 0.15$,  \cr
{\rm non-detection}  & otherwise,                             \cr
}
\label{eq12}
\end{equation}
where $\sigma_{t_{\rm E}}$ is determined from the lightcurve measured 
during the time of $-3\te\leq t-t_0\leq 3\te$.  We note that the same 
detection criteria are imposed on both the first and second events
because determinations of the binary parameters such as the mass ratio 
requires not just simple detections of the companion signature but 
equally well-determined timescales of both events.  We also note that 
the criteria imposed on the current and future surveys are the same 
for impartial comparison of the event rate.  In Figure~\ref{fig:one}, 
we present an example lightcurve of an event that barely passes the 
imposed detection criteria.

\begin{figure}[t]
\epsscale{1.2}
\plotone{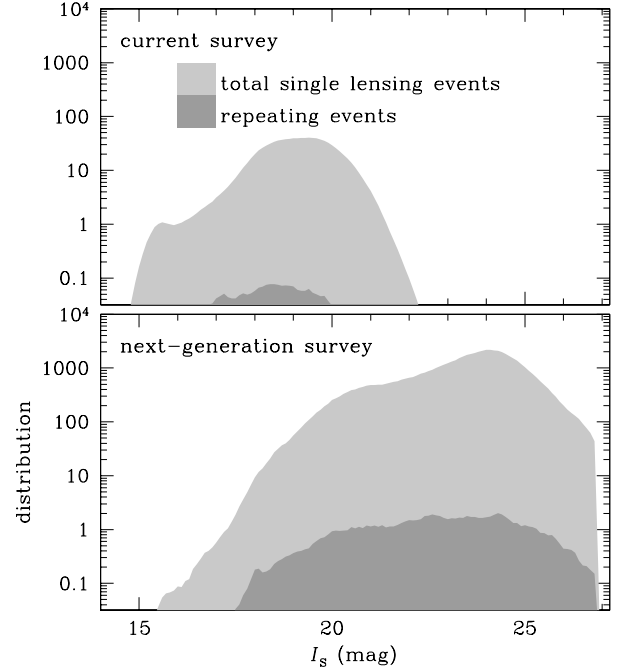}
\caption{\label{fig:two}
Distributions of the source brightness of lensing events to be detectable 
from current and future lensing surveys.  The distributions with light 
greyscale are for the total single lensing events expected when there is 
no binary lenses.  The distributions with dark greyscale are for repeating 
binary events under the binary mass function model where the masses of 
binary components are drawn independently from the same mass function.
Here we assume a binary fraction of $f_{\rm bi}=0.5$.
}\end{figure}

\subsection{Event Rates}

For each event passed the detection criteria, its contribution to the 
total event rate is computed such that 
\begin{equation}
\Gamma_{{\rm p},i} \propto \rho(D_{\rm S}) \rho(D_{\rm L}) 
D_{\rm S}^2 v \Sigma,
\label{eq13}
\end{equation}
where $\rho(D)$ is the matter density along the line of sight, the 
factor $D_{\rm S}^2$ is included to account for the increase of the 
number of source stars with the increase of $D_{\rm S}$, $v$ is the 
lens-source transverse speed, and $\Sigma$ represents the lensing 
cross-section.  The lensing cross-section is proportional to to the 
physical size of the Einstein radius, i.e.\ $\Sigma\propto r_{\rm E}
=D_{\rm L}\thetae$.  Since events can be detected even when the source 
does not enter the Einstein ring,  we set the upper limit of the 
impact parameter large enough to account for the contribution  
of these events to the event rate.  Although our estimation of the
event rate is based on numerical calculation to consider detectability 
depending on the observational conditions, we note that a general 
analytic expression for the rate is found in \citet{distefano99}.

\begin{deluxetable}{lll}
\tablecaption{Comparison of Event Rates \label{table:two}}
\tablewidth{0pt}
\tablehead{
\multicolumn{1}{c}{binary mass} &
\multicolumn{2}{c}{event rate} \\
\multicolumn{1}{c}{function model} &
\multicolumn{1}{c}{current} &
\multicolumn{1}{c}{future}}
\startdata
capture          &  $5.2 f_{\rm bi}/{\rm yr}$ ($1.1 f_{\rm bi}\%$) &  $189 f_{\rm bi}/{\rm yr}$ ($4.3 f_{\rm bi}\%$) \\
fission          &  $1.9 f_{\rm bi}/{\rm yr}$ ($0.4 f_{\rm bi}\%$) &  $91 f_{\rm bi}/{\rm yr}$  ($2.1 f_{\rm bi}\%$) \\
fragmentation    &  $3.0 f_{\rm bi}/{\rm yr}$ ($0.6 f_{\rm bi}\%$) &  $128 f_{\rm bi}/{\rm yr}$ ($2.9 f_{\rm bi}\%$) \\
\enddata
\tablecomments{ 
Comparison of the rates of repeating binary lensing events expected 
from the current and next-generation microlensing surveys under various 
models of binary mass function.  The values in the parentheses are the 
fractions of repeating events out of the total single lensing events 
expected when there is no binary lens.  The rate of total single 
lensing events is $\sim 470$ events/yr for current surveys and 
$\sim 4500$ events/yr for future surveys.
}
\end{deluxetable}

In Table~\ref{table:two}, we summarize the resulting rate of repeating 
binary lensing events expected from the current and next-generation 
microlensing surveys under various binary mass function models.  The 
values in the parentheses are the fractions of repeating events out 
of the total single lensing events expected when there is no binary lens.
We find that the rate of total single lensing events is $\sim 470$ 
events/yr for current surveys and $\sim 4500$ events/yr for future 
surveys.  From the table, one finds that the rate of repeating events 
from the future survey is $\sim 36$--47 times higher than that of the 
current survey depending on the adopted models of the binary mass 
function, reaching $\sim 91f_{\rm bi}$--$189f_{\rm bi}$ per year.  
Considering 3.7-year lifetime of the {\it MPF} mission and assuming 
a binary fraction of $f_{\rm bi} \sim 0.5$ \citep{duquennoy91, fischer92}, 
the total number of binary events expected to be detected via the repeating 
channel would be $\sim 169$--350 events, which is large enough for the 
statistical investigation of the binary properties.

\begin{figure}[t]
\epsscale{1.2}
\plotone{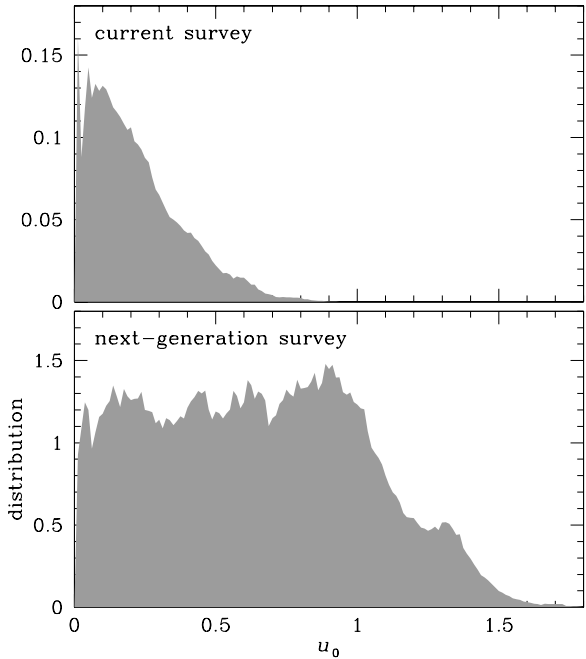}
\caption{\label{fig:three}
Distributions of the impact parameters of events detectable from the 
current (upper panel) and future (lower panel) lensing surveys. 
}\end{figure}

Then, why is the rate of repeating events much higher in future lensing 
surveys.  We find two reasons for this.  The first is the greatly 
enhanced sensitivity of the future survey to events associated with 
faint source stars.  This can be seen in Figure~\ref{fig:two}, where 
we present the distributions of source brightness of the second part of 
the repeating events detectable by the current and future surveys under 
the capture model of the binary mass function.  From the distributions, 
one finds that while the current survey is sensitive only to stars 
brighter than $I\sim 20$, the future survey would be sensitive to stars 
up to $I\sim 26$.  Lensing observation by the current ground-based 
survey is limited  by crowding.  On the other hand, the space-based 
{\it MPF} mission can resolve stars deeper down the luminosity function, 
where there exist much more numerous stars.  As a result, the event 
rate of the future survey would be much higher than the rate of the 
current survey despite the coverage of substantially smaller field of 
view.

Another factor that contributes to the higher detection rate of 
repeating events is the elevated sensitivity to low-magnification 
events.  To demonstrate this, we plot the distributions of the impact 
parameters in Figure~\ref{fig:three}.  From the comparison of the 
distributions, one finds that the average impact parameter of the 
events detectable by the future survey is substantially larger than 
the average value of the events detectable by the current survey, 
implying that the future survey will be sensitive to lower-magnification 
events.  With the increase of the threshold impact parameter, the 
detection efficiency of repeating events also increases.

\begin{figure}[t]
\epsscale{1.2}
\plotone{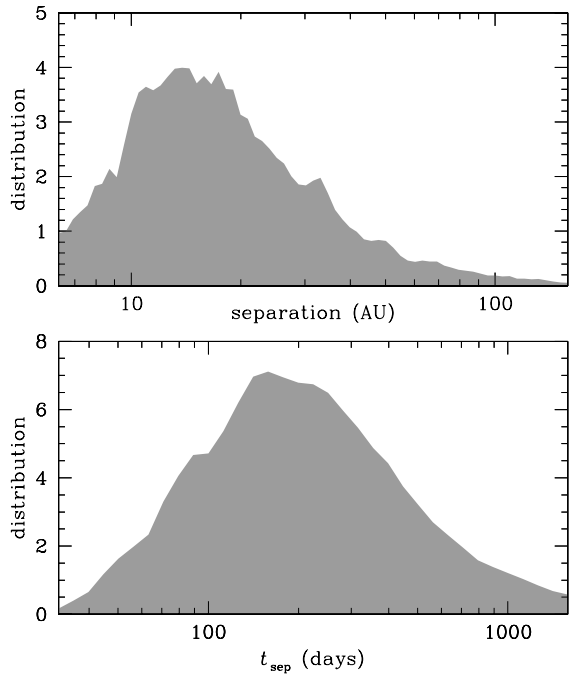}
\caption{\label{fig:four}
Distributions of the binary separations and wait times for repeating 
events to be detected by future lensing surveys.
}\end{figure}

Then, what are the separation range of binaries to be detected through 
the channel of repeating events?  Another important quantity related to 
the observational side is the wait times between the two separate parts 
of repeating events, $t_{\rm sep}$.  In Figure~\ref{fig:four}, we 
present the distributions of the binary separations and wait times.  
We find that the separation distribution peaks at $\sim 15$ AU, but it 
extends considerable separations up to 100 AU.  Therefore, the repeating 
channel would provide sample of binaries with a wide range of separations.
The most typical wait time is $t_{\rm sep}\sim 200$ days and extends up 
to 1000 days.

\section{Discussion}

Binaries can be detected and characterized by various other methods 
including photometric, spectroscopic, and  imaging observations of 
eclipsing, spectroscopic, and visual binaries, respectively.  Studies 
of binaries with microlensing have the following advantages over these 
other methods.  First, while other methods require long-term 
observations, especially for wide-separation binaries, microlensing 
method can detect and characterize binaries from much shorter-term 
observations.  Second, unlike other methods, microlensing method has 
no bias against dark or faint binary components.  Third, while most 
other methods can be applicable only to binaries located relatively at 
small distances from the Earth, binaries to be detected by lensing 
surveys will be located at Galactic-scale distances.  Fourth, 
next-generation lensing surveys will be able to produce a homogeneous 
sample of large binaries.  Therefore, microlensing technique will be 
able to provide better statistics of Galactic population of binaries.

\section{Conclusion}

We explored the usefulness of future lensing surveys in the study of 
binary properties by using the binary sample detectable through a  
channel of repeating events.  To demonstrate this, we estimated the 
rate of repeating events based on standard models of mass function 
and dynamical and physical distributions of Galactic matter combined 
with models of binary separation and companion mass function.  From 
this, we found that the total number of repeating events expected to be 
detected from $\sim 4$-year space-based survey would be $\sim 200$--400, 
which is $\sim 40$--50 times higher than the rate of current surveys, 
and the range of binary separations to be covered by lensing extends 
to up to 100 AU.  We found that the high detection rate is due to the 
greatly improved sensitivity of the future survey to events associated 
with faint source stars and low-magnification events.  Therefore, the 
future lensing surveys will provide a homogeneous binary sample that 
will allow to investigate the statistical properties of Galactic 
binaries unbiased by the brightness of the binary components.

\acknowledgments 

This work was supported by the grant (C00072) of the Korea Research 
Foundation.

\end{document}